\documentstyle[12pt]{article}

\input psfig.sty

\newcommand{\be}{\begin{equation}}
\newcommand{\ee}{\end{equation}}
\newcommand{\bea}{\begin{eqnarray}}
\newcommand{\eea}{\end{eqnarray}}
\newcommand{\gsim}{\lower.7ex\hbox{$\;\stackrel{\textstyle>}{\sim}\;$}}
\newcommand{\lsim}{\lower.7ex\hbox{$\;\stackrel{\textstyle<}{\sim}\;$}}
\newcommand{\tev}{\,{\rm TeV}}
\newcommand{\gev}{\,{\rm GeV}}
\newcommand{\mev}{\,{\rm MeV}}

\newcommand{\ev}{\,{\rm eV}}
\newcommand{\mpl}{M_{\rm pl}}
\newcommand{\mst}{M_{*}}

\newcommand{\nwall}{N_{\rm wall}}

\begin{document}

\baselineskip=17pt
\pagestyle{plain}
\setcounter{page}{1}

\begin{titlepage}

\begin{flushright}
SLAC-PUB-7949\\
CERN-TH/98-297\\
SU-ITP-98/54

\end{flushright}
\vspace{5 mm}

\begin{center}
{\LARGE Stabilization of Sub-Millimeter Dimensions:}
\vskip 2mm
{\LARGE  The New Guise of the Hierarchy Problem}
\vspace{3mm}
\end{center}
\begin{center}
{\large Nima Arkani-Hamed$^a$, Savas Dimopoulos$^b$ and
John March-Russell$^c$}\\
\vspace{2mm}
{\em $^a$ SLAC, Stanford University, Stanford CA 94309,  
USA}\\
{\em $^b$ Physics Department, Stanford University, Stanford CA 94305,  
USA}\\
{\em $^c$ Theory Division, CERN, CH-1211, Geneva 23, Switzerland}
\end{center}
\vspace{2mm}
\begin{center}
{\large Abstract}
\end{center}
\noindent
A new framework for solving the hierarchy problem was recently  
proposed which does not rely on low energy supersymmetry or  
technicolor.  The fundamental Planck mass is at a $\tev$ and
the observed weakness of gravity at long  
distances is due the existence of new sub-millimeter spatial  
dimensions.  In this picture the standard model fields are
localized to a $(3+1)$-dimensional wall or ``3-brane''.
The hierarchy problem becomes isomorphic to the  
problem of the largeness of the extra dimensions.  This is in
turn inextricably linked to the cosmological constant problem,
suggesting the possibility of a common solution.  The radii of the
extra dimensions must be prevented from both expanding to too
great a size, and collapsing to the fundamental Planck length
$\tev^{-1}$.  In this paper we propose a number of mechanisms
addressing this question.  We argue that a positive bulk
cosmological constant $\bar\Lambda$ can stabilize the internal
manifold against expansion, and that the value of $\bar\Lambda$ is not
unstable to radiative corrections provided that the supersymmetries
of string theory are broken by dynamics on our 3-brane.
We further argue that the extra dimensions can be stabilized against
collapse in a phenomenologically successful way by either
of two methods:  1) Large, topologically conserved quantum numbers 
associated with higher-form bulk U(1) gauge fields, such as the
naturally occurring Ramond-Ramond gauge fields, or the winding
number of bulk scalar fields.  2) The brane-lattice-crystallization
of a large number of 3-branes in the bulk.  These mechanisms are
consistent with theoretical, laboratory, and cosmological
considerations such as the absence of large time  
variations in Newton's constant during and after primordial
nucleosynthesis, and millimeter-scale tests of gravity.
\end{titlepage}
\newpage


\section{New Guise of the Hierarchy Problem}

A new proposal for solving the hierarchy problem was 
recently introduced \cite{ADD,AADD,ADDlong} which circumvents
the need for supersymmetry or technicolor.  Instead the
hierarchy problem for the 
standard model (SM) is solved by bringing the fundamental Planck scale
down to the $\tev$ scale.  Gravity becomes comparable in
strength to the other interactions at this scale, and the
observed weakness of gravity at long distances is then explained
by the presence of $n$ new ``large" spatial dimensions. 

Gauss' Law relates the Planck scales of the
$(4+n)$-dimensional theory, $\mst$, and the long-distance
4-dimensional theory, $\mpl$,
\be
\mpl^2 \sim r_n^n \mst^{n+2}
\ee
where $r_n$ is the size of the extra dimensions.
Putting $\mst \sim 1\tev$ then yields
\be
r_n \sim 10^{-17+\frac{30}{n}} \mbox{cm}
\ee
For $n=1$, $r_1 \sim 10^{13}$ cm, so this case is
excluded since it would modify Newtonian gravity at
solar-system distances. Already for $n=2$, however, $r_2 \sim 1$
mm, which happens to be the distance where our present experimental
knowledge of gravitational strength forces ends.  For larger
$n$, $1/r_n$ slowly approaches the fundamental Planck scale
$\mst$. 

While the gravitational force has not been measured beneath
a millimeter, the success of the SM up to $\sim 100\gev$ implies
that the SM fields can not feel these extra large dimensions; that
is, they must be stuck on a 3-dimensional wall, or ``3-brane", in the  
higher dimensional space.  Thus, in this framework the universe is  
$(4+n)$-dimensional with fundamental Planck scale near the weak
scale, with $n \geq 2$ new sub-mm sized dimensions where gravity,
and perhaps other fields, can freely propagate, but where the SM
particles are localised on a 3-brane in the higher-dimensional space.
The most attractive possibility for localizing the SM fields to
the brane is to employ the D-branes that naturally occur in type I
or type II string theory \cite{Dbrane,AADD}.  Gauge and other
degrees of freedom are naturally confined to such
D-branes~\cite{Dbrane}, and furthermore this approach has the obvious
advantage of being formulated within a consistent theory of gravity.
However, from a practical point of view, the most important question
is whether this framework is experimentally excluded.  This was the
subject of \cite{ADDlong} where laboratory, astrophysical, and
cosmological constraints were studied and found not to exclude these
ideas.

There are also a number of other important papers discussing related
suggestions.
Refs.~\cite{DDG} examine the idea of lowering the GUT scale by
utilizing higher dimensions.  Further papers
concern themselves with the construction of string models with extra  
dimensions larger than the string scale \cite{antoniadis,HW,tye},
and gauge coupling unification in higher dimensions
without lowering the unification scale~\cite{bachas}.
There are also two important papers by Sundrum.  The first deals
with the effective theory of the low energy degrees of freedom in
realizations of our world as a
brane~\cite{Raman1}, while the second is concerned with
the topic of radius stabilization~\cite{Raman2}, and with which our
analysis has much in common.

In our framework the hierarchy problem becomes the problem of  
explaining the size and stability of the large extra dimensions.  
The main purpose of this paper is to exhibit mechanisms which   
accomplish these objectives, and examine some aspects of their
phenomenology.  Since a rather wide collection of possible
stabilization mechanisms are discussed in this paper, only
some of which we believe to be successful, we think it
useful to provide the reader with a guide to our main results:
In Section~1.1 we discuss a very general consistency
constraint on the bulk cosmological constant; and in Section~2
we describe some basic kinematics pertaining to the radial
oscillation field, whose mass will turn out to provide significant
constraints on stabilization scenarios.  In particular this is the
constraint that will sometimes force us to have a large conserved
integer parameter in our models.  In Section~3 we show that the
properties and limits on such light radial oscillation fields can be
discussed in a way that is independent of the details of the
precise radius-stabilization mechanism.  We also briefly describe
the reasons for the cosmological safety of this scenario.
The most important
results of this paper are contained in Section~4 where we discuss
long-distance (IR) and, particularly, short-distance (UV) stabilization
mechanisms, and put these together to obtain a variety of complete 
stabilization models.  We find that two methods of UV stabilization
are particularly attractive: ``brane-lattice-crystallization'' discussed
in Section~4.2; and ``topological stabilization'' discussed
in Section~4.3.  Finally in Section~5 we present a summary
of our results.

\subsection{The Hierarchy and the Bulk Cosmological Constant.}

Let us begin with some necessary conditions that must be satisfied to  
ensure the existence of large radii. As we know from experience with  
our 4-dimensional world, to ensure that our three ordinary spatial  
dimensions are very large the radius of curvature of the universe
must be no less than the present horizon size. This leads to  
the requirement that the  cosmological constant of the universe is less  
than the critical density.  An identical line of reasoning for the case  
of $n$-extra dimensions also leads to an upper limit on the bulk  
cosmological constant as we now explain~\cite{ADMR}.

The curvature radius $L_{\rm curv}$ of the bulk
space in the presence of energy density or an effective cosmological
constant, $\bar \Lambda$, in the bulk, is
\be
L_{\rm curv} \sim \left( {\mst^{n+2}\over \bar \Lambda}\right)^{1/2} .
\label{Lcurv}
\ee
This curvature
radius must be larger than the physical size of the transverse
dimensions $r_n$ in order to insure  that the bulk space does not  
``split off'' into separate inflating universes separated by horizons
of size $L_{\rm curv}$, or collapse into black holes. This  
gives an upper bound on $\bar \Lambda$~\cite{ADMR}:
\be
\bar \Lambda\lsim \mst^{(4+n)}
\left( {\mst\over \mpl}\right)^{4/n}
\label{lambdabound}
\ee
This constraint will play an important role in what follows. It already  
implies that the magnitude $\bar \Lambda$ must be smaller than the  
fundamental scale of $M_*$.  This was to be expected since in this case  
there is one scale in the problem and the bulk would split into a  
collection of non-communicating $1/\tev$ size regions, outside of each
others' particle horizons.
An important corollary of this is that one cannot use the  
Scherk-Schwarz mechanism to break supersymmetry at $M_*$ since this  
would induce a bulk cosmological constant of the order of $M_*^{4+n}$,
which exceeds the limit Eq.~(\ref{lambdabound}).

Of course the effective 4-dimensional cosmological constant
measured at long distances (greater than the size of the extra
dimensions) must to a very high degree of accuracy vanish.  This
can be achieved by cancelling the wall and bulk contributions
against on another:
\be
0 = f^4 + (r_n)^n \bar\Lambda
\label{IRcosmo}
\ee
We see that if the bulk energy is negative, a positive $f^4$ will
cancel the 4-dimensional cosmological constant,
while if the bulk energy is positive, we need a negative $f^4$. Clearly
a positive $f^4$ is reasonable; if the wall 
can fluctuate in the extra dimensions, $f^4$ is just the tension
of the wall, and provides the correct sign kinetic 
term for the Nambu-Goldstones of spontaneously broken $(4+n)$-dimensional
Poincare invariance which live on the wall. 
This reasoning seems to exclude the possibility of a negative
$f^4$, since this gives the wrong sign kinetic term to the Nambu-Goldstones.
This is however only a problem if the Nambu-Goldstone fields are
indeed present, that is, if the $(4+n)$-dimensional Poincare invariance 
is {\it spontaneously} broken. On the other hand, suppose that the wall
is ``stuck" and cannot fluctuate in the extra dimensions, due to 
{\it explicit} breaking of $(4+n)$-dimensional Poincare invariance.
As an example, we can consider twisted sector fields living at an
orbifold fixed point.  In the language of string theory the wall
could be an orientifold rather than a D-brane.
In this case, $f^4$ is just the wall energy density acting as a
source for gravity, but there are no Nambu-Goldstones on the wall to
receive a wrong-sign kinetic term.
Another way of saying this is as follows.  The wall can have an
energy density as a source for gravity $f^4_{\rm grav}$,
and a tension under ``bending'' $f^4_{\rm bend}$.
It is $f_{\rm grav}^4$ which should appear in Eq.(\ref{IRcosmo}).
If the $(4+n)$-dimensional Poincare invariance is only spontaneously
broken, its non-linear realization forces 
$f^4_{\rm grav}=f^4_{\rm bend}$, as they both come from expanding
the term in the action
\be
-\int d^4 \sqrt{-g_{ind}} f^4 ,
\ee
where $g_{ind}$ is the induced metric on the wall.  Since
$f^4_{\rm bend}>0$, we have $f^4_{\rm grav}>0$.  On the other hand,
if the $(4+n)$-dimensional Poincare invariance is explicitly broken,
there need not be any 
relationship between the two.  Indeed, if the wall can not fluctuate,
effectively $f^4_{\rm bend} = \infty$, while $f^4_{\rm grav}$ can
be finite and of any sign.  

We will therefore allow the possibility that a brane can make a net
negative contribution to the 4-dimensional cosmological constant, which
provides us with the freedom to consider stabilization mechanisms that
give either positive or negative bulk energy densities.\footnote{We
thank Eva Silverstein for discussions about this point.}

Given Eq.~(\ref{lambdabound}) we learn that if our wall is the only
brane, then its effective
wall-localized cosmological constant, $f^4$, is bounded above by
\be
f \lsim \mst \left( {\mpl\over \mst}\right)^{(n-2)/2n}
\label{wallcosmo}
\ee
This is not too severe a constraint though, varying between
$10\tev$ for $n=2$, to $\sim10^8 \gev$ for $n=6$.  Of course,
the relation, Eq.~(\ref{IRcosmo}), can be turned around to determine
the effective bulk cosmological constant, $\bar\Lambda$,
given $f$.  A natural assumption for the wall-localized cosmological
constant, given our state of knowledge of the standard model
interactions on the wall, is $f^4 = (1\tev)^4 \sim \mst^4$.  Thus
in this case
\be
\bar \Lambda = \mst^{4+n}\left({\mst\over \mpl}\right)^2
\label{bulkcosmo}
\ee
is the value of the bulk cosmological constant necessary to cancel
the total long-distance cosmological constant in our world. 
Note that this value is indeed always less than the upper bound
Eq.~(\ref{lambdabound}) arising from the bulk curvature constraint.

Later in Section~4.2 we will consider stabilization mechanisms
that utilize many branes populating the bulk.  In this case the 
bounds Eqs.~(\ref{lambdabound}) and (\ref{wallcosmo}) are modified
by the total brane number $\nwall$. 

A difference with the previous case is that the curvature
radius must now only be greater than the inter-brane separation
$r_n/(\nwall)^{1/n}$. (We are assuming the best case situation 
of equally spaced branes which leads to the weakest bound.)
The reason for this is that the branes themselves are localized
sources of curvature of the opposite sign, so that at long distances
compared to the inter-brane separation, the curvature of the bulk
averages out to zero.
From this follows the generalized curvature constraint
\be
\bar \Lambda\lsim \nwall \mst^{(4+n)}
\left( {\mst\over \mpl}\right)^{4/n} .
\label{gencurvconst}
\ee
The IR cancellation of the
effective cosmological constant in 4-dimensions is expressed by
\be
0 = \nwall f^4 + (r_n)^n \bar\Lambda .
\label{IRcosmoN}
\ee
Imposing this leads to the following bound on the wall-localized
cosmological constant
\be
f^4 \lsim \mst^4 \frac{1}{\nwall^{(n-2)/n}}
\left(\frac{\mpl}{\mst}\right)^{(2n-4)/n}.
\label{wallcosmoN}
\ee 

The cosmological constant is bounded from below from another  
consideration.  As we will see later in Section~3,
there are light gravitationally coupled  
particles in the spectrum whose $({\rm mass})^2$ is proportional to
$\bar\Lambda$ (see Eq.~(\ref{rmass})).
The requirement that these particles do not conflict with  
measurements of gravity imply that they weigh more than a meV and  
consequently put a lower limit on $\bar \Lambda$.  This in turn implies   
that the large size of the new dimensions in most, but not all cases  
studied here, cannot be solely due to the smallness of $\bar \Lambda$.
Additional dynamics to boost the size of the extra dimensions are
necessary.  This can easily come about if there is a conserved charge
in the system, analogous to baryon number.  Just as humans are
large because they carry large  
baryon number, the extra dimensions can be large because they carry  
some large charge $Q$.   In some of our examples, this charge
corresponds to a large number of walls 
$Q \sim N_{\rm wall} \gg 1$. In others, it is a topological charge $k$. 
Note however that in some special cases, it is not necessary to use
a large conserved charge.  For example, as we discuss in Section~4.3 if
the fundamental scale $M_*$ is pushed to $\sim 10\tev$ while the
wall contribution to the cosmological constant $f^4$ is kept at
$\sim (1\tev)^4$, then topological charges $k\sim 1$ are adequate.
This is not too unnatural a situation, especially
considering that a loop factor could easily supply such
a suppression to $f^4$.

\subsection{Stable and Calculable Hierarchy}

In this paper we will not search for dynamical  
mechanisms where the hierarchy between the size of the extra dimensions  
and the fundamental scale is {\it calculable}. We will instead be content to  
enforce this hierarchy by choosing the bulk cosmological constant to be  
small and/or the above-mentioned topological or other charge to be  
large. This is analogous to the early days of the minimal supersymmetric  
standard model (MSSM)~\cite{dimopoulosgeorgi81} where the soft  
supersymmetry breaking terms were postulated without any reference to a  
dynamical mechanism which generates them. The idea there was that since  
the problem of supersymmetry breaking is connected with the  
cosmological constant problem it seemed premature to adopt a specific  
SUSY-breaking mechanism and it seemed more prudent to study  
consequences that were independent of the details of the SUSY-breaking  
mechanics.  Similarly,
in our new framework the hierarchy and cosmological constant problems  
are even more closely intertwined so we will adopt a similar philosophy  
of not insisting on a detailed dynamical mechanism for a {\it calculable}  
hierarchy and will be content to instead parametrize our ignorance by a  
choice of $\bar \Lambda$ and an integer $Q$ ($\nwall$ or $k$).

The second aspect of the hierarchy problem is its stability against  
radiative corrections. In the MSSM this is guaranteed by  low energy  
supersymmetry, which protects the Higgs mass against large radiative  
corrections. 
Presumably, the analogous question in our framework is the behaviour of   
the pair of parameters ($\bar \Lambda, Q$) under radiative corrections.  
The integer $Q$ is automatically protected since it refers to charge of a  
configuration. Since  $\bar \Lambda$  is a bulk cosmological constant  
one can imagine two possibilities. One is that whatever solves the  
cosmological constant problem will also prevent $\bar \Lambda$ from  
becoming as large as the cutoff $\mst$. The second more explicit and  
perhaps more satisfactory viewpoint is to invoke bulk-supersymmetry to  
protect $\bar \Lambda$ from large radiative corrections. Indeed, as  
pointed out in reference \cite{AADD}, if supersymmetry is broken solely  
on our 3-brane by an amount $\sim M_* \sim 1\tev$, the Fermi-Bose  
splittings that this induces in the bulk are miniscule
$ \sim {\tev^2/\mpl} \sim 10^{-3}\ev$ and therefore the bulk
cosmological constant 
$\bar \Lambda$ is protected by the approximate bulk-supersymmetry.

It should be emphasized that stabilizing {\it large} dimensions is
inherently easier than stabilizing Planck-scale dimensions.  In the
latter case, quantum gravitational effects are necessarily important and 
can not be ignored.  However, precisely because we are interested in
large radii, the details of short distance physics are largely irrelevant
and a classical or semi-classical analysis suffices.  We will consider
this point more explicitly in Section~4.

\section{Kinematics of Radius Stabilization}

Suppose that we have an $N$-brane embedded in a space with 
$N$ large spatial dimensions and $n$ small
dimensions we wish to stabilize.  The total action is comprised
of a bulk part,
\be
S_{\rm bulk}  = - \int d^{1+N+n}x
\sqrt{-\det G_{(1+N+n)}} \biggl( \mst^{(n+N-1)}{\cal R} +\Lambda
- {\cal L}_{\rm matter} + \ldots \biggr), 
\label{bulkS}
\ee
and a brane part,
\be
S_{\rm brane}  =  - \int d^{1+N}x \sqrt{-\det g_{(1+N)}^{\rm induced}}
\biggl( f^{N+1} + \ldots \biggr),
\label{actions}
\ee
where ${\cal L}_{\rm matter}$ is the Lagrangian of bulk gauge or
scalar fields, and the ellipses denote higher-derivative terms
that can be ignored in the regime of interest as we will demonstrate
below.   Take the background metric for the $(1+N+n)$-dimensional
spacetime to be of the form
\be
g_{\mu\nu} = \left(
\begin{array}{ccc}
1 & & \\
   & - R(t)^2 g_{IJ} & \\
        & & -r(t)^2 g_{ij}
\end{array}
\right) ,
\label{metricform}
\ee
where $R$ is the scale factor of the $N$-dimensional space, and $r$ is
the scale factor of the internal $n$-dimensional space, with geometry
set by $g_{ij}$ where $\det(g_{ij}) = 1$.

With this metric the Ricci scalar is 
\be
-{\cal R} = 2N{ \ddot R\over R } + N(N-1)\left({ \dot R\over R }\right)^2
+ 2n{ \ddot r\over r } + n(n-1)\left({ \dot r\over r }\right)^2 +
2Nn\left({ \dot r\dot R\over rR }\right) + {\kappa n(n-1) \over r^2},
\label{ricci}
\ee
where the internal curvature term is present for $n$-spheres
($\kappa=1$), but vanishes for tori ($\kappa=0$), and we have ignored
a similar curvature term for the large dimensions.  After integrating
over all spatial coordinates we obtain,
\be
S = \int dt \bigl( {\cal L}_{\rm KE}(\dot R, \dot r) - R^N V_{\rm tot}(r)
\bigr),
\label{intaction}
\ee
where the total potential is given by
\begin{eqnarray}
V_{\rm tot}(r)  &=& V_{\rm bulk} + V_{\rm wall} \nonumber \\
V_{\rm wall} &=& f^{N+1}  \nonumber \\
V_{\rm bulk} &=& \Lambda r^n - n(n-1)\kappa \mst^{n+N-1} r^{n-2}
 + V_{\rm matter}(r)
\label{potentials}
\end{eqnarray}
where,
\be
V_{\rm matter}(r)  =  - \int d^n x \,  \bigl(r^n {\cal L}_{\rm matter}
\bigr) .
\ee
After integrating the $\ddot R$ and $\ddot r$ terms by parts, the kinetic
part of the action for the radii, $R$ and $r$, becomes
\be
S = -\mst^{N+n-1}\int dt\, R^N r^n
\left( N(N-1)\left({ \dot R\over R }\right)^2
+  n(n-1)\left({ \dot r\over r }\right)^2 +
2Nn\left({ \dot r\dot R\over rR }\right) \right) .
\label{radiiKE}
\ee
Note the overall negative sign of these kinetic terms.
This is connected to the well-known phenomenon
that the conformal mode of gravity has the opposite sign kinetic
term to the transverse graviton kinetic term (and which bedevils
attempts at defining quantum gravity via the Euclidean functional
integral).

In any case there is clearly an extremum of the action with $\dot R=\dot r=0$,
when the condition $\partial_R ( R^N V_{\rm tot}(r))|_{R=R_0, r=r_0}=0$,
(and similar with $\partial_R\to\partial_r$) is met.  These imply (for
$R_0\neq0$)
\bea
V_{\rm tot}(r_0) & = & 0, \qquad {\rm and}\cr
V'_{\rm tot}(r_0) &= & 0 .
\label{staticcond}
\eea
This is as one would have naively expected.
However, because of the negative sign for the kinetic term for the radial
degrees of freedom, the stability analysis for such static solutions has
to be treated with care.  The analysis starts by expanding the action,
Eq.~(\ref{radiiKE}), in small fluctuations around the extremum:
$R(t) = R_0 +\delta R(t)$, and $r(t) = r_0 +\delta r(t)$.  
Then to quadratic order, and defining $\Delta \equiv \delta R/R_0$ and
$\delta \equiv \delta r/r_0$, the expansion gives the coupled equations
of motion
\be
\left(
\begin{array}{cc}
N(N-1) & Nn \\
Nn & n(n-1) 
\end{array}\right)
\left(
\begin{array}{c}
\ddot \Delta \\
\ddot \delta
\end{array}\right)
= 
\left(
\begin{array}{cc}
0 & 0 \\
0 & \omega^2 
\end{array}\right)
\left(
\begin{array}{c}
\Delta \\
\delta
\end{array}\right),
\label{stabM}
\ee
where 
\be
\omega^2 = \frac{1}{2} { (r_0)^2 V''_{\rm tot}(r_0)\over
\mst^{N+n-1}(r_0)^n} = \frac{1}{2} { (r_0)^2 V''_{\rm tot}(r_0)\over
M_{(N+1)}^{N-1}}.
\label{omegadef}
\ee
Here $M_{(N+1)}$ is to be understood as the effective Planck
mass in the large $(N+1)$-dimensional spacetime
($M_{(4)}\equiv\mpl$).   We now search
for oscillating solutions, $(\Delta,\delta) =
\exp(i\Omega t)(\Delta_0,\delta_0)$ of the stability equations.
From Eq.~(\ref{stabM}), $\Omega^2$ is thus given by the
eigenvalues of the matrix
\be
+{\omega^2\over nN(N+n-1)}
\left(
\begin{array}{cc}
0 & - Nn \\
0 & N(N-1) 
\end{array}\right),
\label{Omegaeigen}
\ee
namely, $\Omega^2 =0$, and 
\be
\Omega^2 = {(N-1)\over n(N+n-1)}\omega^2.
\label{Omegaeqn}
\ee

The zero eigenvalue just corresponds to the fact that $R_0$ is
a flat direction since, by assumption, there is no potential
for $R$.  The crucial expression is Eq.~(\ref{Omegaeqn}), which
gives us the condition for stability of our static solution.
Stability requires $\Omega^2 >0$, which for $N>1$ implies
\be
\omega^2 >0 \quad \Rightarrow V''_{\rm tot}(r_0) >0 .
\label{stability}
\ee
This is the main result of this Section.  Even though it seems
trivial that stability is equivalent to requiring the second
derivative of the potential around the extremum to be positive,
this condition is {\it a priori} not at all obvious given 
the negative kinetic terms for the radii fields.  As an example
of this consider the case $N=0$, which corresponds to $r$ being
thought of as the radius of a Friedman-Robertson-Walker universe.
{\it In this case} stability requires $\omega^2 < 0$, or equivalently
$V''_{\rm tot}(r_0) < 0$.  This accords with our usual understanding:
for example take the only term in $V$ to be a positive cosmological
term $V_{\rm tot}(r) = \Lambda r^n$.  Then around the minimum
at $r=0$ the solution is unstable to inflationary growth as we
expect. 

The end result of this analysis is simply that we can think in terms
of a total potential $V(r)$ that one can minimize to find the stable
static solutions for the size of the internal dimensions.
Also note that from Eqs.~(\ref{Omegaeqn}) and (\ref{omegadef}) we 
can extract
the mass of the canonically normalized radial oscillation field $\phi$
(we will refer to $\phi$ as the ``radion'') in the case of interest,
$N=3$, $n$ arbitrary:
\be
m^2_{\rm radial} = \frac{1}{n(n+2)}
{(r_0)^2 V''_{\rm tot}(r_0) \over \mpl^2 }
\label{rmassnorm}
\ee
Notice that as a consequence, the magnitude of $\phi$ is related
to the deviation $\delta r$
from the equilibrium radius $r = r_0 + \delta r$ via
\be
\frac{\delta r}{r_0} \sim \frac{\phi}{M_{\rm pl}} .
\ee

\section{Model-independent limits on light radions}

Before we move on to the very important issue of the explicit 
nature of possible radius stabilization mechanisms as discussed
in Section~4, it will be
useful for us to examine some model-independent features 
of all these mechanisms.  These include
the existence of a light radial oscillation field
$\phi$, with known couplings to standard model fields.
Although such a field seems to be dangerous, we will argue
below that it satisfies the various limits.

To see that independent of the details of the 
stabilizing potential there is an upper bound
on the mass of the radial excitation field it is useful
to consider a general form for the
bulk stabilizing potential $V_{\rm bulk}(r)$.
{\it Around the equilibrium position} this potential
can be well approximated by the sum of just two
powers of $r$:\footnote{In this paper we will not explicitly consider
potentials of the form $V_{\rm bulk}(r) \sim r^\alpha f[\log(r)]$,
for some function $f$, with only a single such term
dominantly contributing to the potential energy near the
equilibrium position.  See Ref.~\cite{ADDlong} for a discussion
of such potentials.}  
\be
V_{\rm bulk} (r) = M_*^4 \left(A x^a + B x^b\right).
\ee
Here we have introduced the dimensionless radius variable
$x \equiv r M_*$. 
In particular, following on from the discussion in Section~2, for
a stable minimum we study potentials of the form
\be
V_{\rm bulk}(r) = M_*^4 \left(\epsilon x^\alpha +
\frac{N}{x^\beta} \right), \qquad \alpha,\beta>0 ,
\label{pot1}
\ee 
or
\be
V_{\rm bulk}(r) = M_*^4 \left(\epsilon x^\alpha -
\eta x^\beta \right), \qquad \alpha>\beta>0.
\label{pot2}
\ee
As we discuss in Section~4 the dimensionless
parameter $\epsilon$ is a measure of the size of the
effective bulk cosmological constant, and acts
to prevent the radius from expanding to infinity.
In contrast, the $N$ or $\eta$ terms prevent collapse
to the UV, and arise from either inter-brane interactions,
or from the kinetic energy of topologically quantized bulk
gauge or scalar fields.  
As we will soon see, to get a large radius requires 
a small $\epsilon$, and/or a large $N$ or $\eta$.

Requiring the cancellation of the effective 4-dimensional cosmological
constant at the minimum of these potentials leads to the
equations
\bea
V'_{\rm bulk}(r_0) &=& 0, \cr
V_{\rm bulk}(r_0) + \nwall \bar{f}^4 &=& 0.
\label{minimize}
\eea
Here we have allowed for the possibility that there is
more than one wall or brane in the bulk, $N_{\rm wall}\geq 1$. 
These provide localized sources of curvature (in principle
of either sign as discussed in Section~1.1).  However, for
simplicity, we have assumed that all the branes have broadly similar
such energy densities $f_{{\rm grav},i}^4 \simeq \bar{f}^4$, and
of the same sign.  More general possibilities can also be analyzed.

In any case, the equations~(\ref{minimize}) can be used to determine
$r_0$ and the required value of $N_{\rm wall} \bar{f}^4$ in terms of
the basic model-dependent parameters of the potential,
$\alpha,\beta, \epsilon$, etc.  Alternately, we can find the
values of these parameters necessary to produce a desired internal
radius $x_0 =r_0\mst$ (in string units).  Defining the useful
dimensionless combination 
\be
\gamma \equiv N_{\rm wall} \frac{\bar{f}^4}{\mst^4}
\label{gammadef}
\ee
the stabilizing parameters $\epsilon$ and
$N$ are determined to be:
\be
\left.
\begin{array}{ccc}
\epsilon &=& -\frac{\beta}{\alpha + \beta}\gamma
\frac{1}{x_0^\alpha}\cr
&&\cr
N&=& -\frac{\alpha}{\alpha + \beta}\gamma
x_0^\beta \cr
\end{array}\right\}
~{\rm given}~ \bar{f}^4  <0 .
\label{Neqn}
\ee
In the case of the
potential~Eq.~(\ref{pot2}),
\be
\left.
\begin{array}{ccc}
\epsilon &=& \frac{\beta}{\alpha - \beta}\gamma\frac{1}{x_0^\alpha},\cr
&&\cr
\eta &=& \frac{\alpha}{\alpha - \beta}\gamma\frac{1}{x_0^\beta},\cr
\end{array}\right\}
~{\rm given}~ \bar{f}^4  >0 . 
\label{etaeqn}
\ee

Now, by equipartition, the second derivative of the general potential
$V(r)_{\rm bulk}$ of Eqs.~(\ref{pot1}) and (\ref{pot2}) around
the minimum is given
by $V'' \sim V(r_0)_{\rm bulk}/(r_0)^{2}$.  In addition
the mean bulk value of the cosmological constant is defined by
\be
\bar \Lambda \equiv {V(r_0) \over (r_0)^n}.
\label{lambdadef}
\ee
Thus using the definition
of the canonically normalized radial excitation, Eq.~(\ref{rmassnorm}),
it is easy to see that physical mass of the radial excitations is
\be
m^2_{\rm radial} \sim {\bar \Lambda \over \mst^{2+n}}.
\label{rmass}
\ee
But now we can apply the curvature radius bound on $\bar\Lambda$,
Eq.~(\ref{lambdabound}), to find
\be
m^2_{\rm radial} \lsim \mst^2 \left({ \mst \over \mpl}\right)^{4/n} 
\nwall \sim
\frac{\nwall}{r_n^2}
\label{rmassbound}
\ee
{\it independent} (up to the ${\cal O}(1)$ coefficients we have
dropped) of any details of the stabilizing potential or mechanism.
Evaluating this for the most conservative case of
$\nwall =1$ and for the desired values of $\mst$ leads to a mass
for the radial field that varies between $10^{-2}\ev$ or less for  
$n=2$,
to $\sim 20\mev$ or less for $n=6$.  Note that the reason why the radion  
mass is much smaller than $\mst$ is that $\bar \Lambda$ must be
relatively small to allow large extra dimensions.  

So, in all the models for radius stabilization that we consider,
the radion field will be very light 
with $m^2_{\rm \phi} \lsim r_{n}^{-2}$, at most $\sim 20\mev$ for $n=6$.
Thus it is necessary to study the model-independent limits on such
light radions to make sure that the entire scenario is not excluded.
To do this, we have to determine the coupling of radions to SM fields
on our wall.  At first, it seems that there is no direct coupling of
the radion field to SM fields.  The reason is that the couplings of
SM fields to gravity all come from the induced metric on our wall,
which, if the possible Nambu-Goldstones on the wall are turned off so the
wall is flat in the extra dimensions, depends on $g_{\mu \nu}$ but
not the radion fields $g_{mn}$.   However, this argument is not 
correct and the radion field does indeed couple to SM fields as 
we now show.\footnote{We thank Raman Sundrum and Riccardo Rattazzi for 
setting us straight on this point.} 
Let us go to the effective theory at distances large compared to the size 
of the extra dimensions.  The effective action is
\be
\int d^4 x \sqrt{-{\rm det} g_{\mu \nu}} \left( - M_{\rm pl}^2 \left\{ 1 + 
\frac{n\phi}{M_{\rm pl}} +
\cdots\right\} {\cal R} + g^{\mu \nu} \partial_{\mu}\phi 
 \partial_{\nu} \phi - m^2_{\phi} \phi^2 + {\cal L}_{SM}(\psi,g_{\mu
\nu}) \right),
\ee
where $\psi$ are the SM fields.  Notice that since the effective
4-dimensional Planck scale depends on the size of the
extra dimensions, there is $\phi$ dependence in the coefficient of ${\cal R}$,
and that there is no explicit dependence on $\phi$ in the SM part of the 
Lagrangian as expected.  However, there is kinetic mixing between
ordinary gravity and the $\phi$ field, specifically if we expand
around a flat metric $g_{\mu \nu} = \eta_{\mu \nu}
+ h_{\mu \nu}$, there is a mixing of the form $\phi \partial^2 h^{\mu}_{\mu}$.
Thus, even though there is no direct coupling of the SM fields to $\phi$,
one is induced through this mixing.  This can be seen more clearly if we 
first perform a Weyl rescaling to remove the $\phi$ dependence in front of 
the usual graviton kinetic term. The coupling to SM fields then comes from the 
scale-invariance violating part of the SM lagrangian; the leading interaction
is 
\be
{\cal L}_{SM-\phi} = \frac{\phi}{M_{\rm pl}} (T_{SM})_{\mu}^{\mu} .
\ee
Note that $\phi$ interactions are suppressed for relativistic particles, 
while it has comparable strength to gravity for non-relativistic 
particles. Suppose that $\phi$ is massless.
As far as the long-range force between non-relativistic
particles is concerned, this just amounts to a redefinition of Newton's
constant $G_N \to G_{N, {\rm non-rel.}}$, while the Newton constant governing 
the interaction of gravity with light $G_{N, {\rm rel.}}$ retains the 
standard value $G_{N, {\rm rel.}}=G_N$.  However, the successful predictions
for the gravitational deflection of light as well as Big-Bang Nucleosynthesis
assume $G_{N, {\rm non-rel.}} = G_{N, {\rm rel.}}$
at least to within a few percent.  Moreover,
since the long range force between non-relativistic masses has been 
measured down to $\sim 1$~mm without revealing any deviation from 
Newtonian gravity, the mass 
of $\phi$ must be pushed up above $\sim (1 {\rm mm})^{-1} \sim 10^{-3}\ev$.
This model-independent constraint, which was also discussed in 
\cite{Raman2}, is the most important limit on light radions.

We now move on to consider other possible limits on light radions   
coming from cosmology and astrophysics.  There are two classes of 
worries.  The first is that not only the radion but all of its Kaluza-Klein
excitations can be produced in the early universe and in stars, leading
to such well-known problems as the over-efficient cooling of supernovae.
This concern is identical to the problem of bulk graviton overproduction 
which was studied in \cite{ADDlong}, and found to in some cases (namely
$n=2$ extra dimensions) to constrain but not rule out our scenario.
The second concerns oscillations of the radion 
field itself around its minimum.  These may overclose the universe, and
further, since these oscillations correspond to changing 
the size of the extra dimensions, they also lead to an oscillating
4-dimensional Newton's constant, which can be problematic.\footnote{We
remark that {\it high-frequency} oscillations of $G_N$ of sufficiently
small amplitude around a mean value equal to the standard value of
$G_N$ can be accommodated, despite the fact that in such a case $dG_N/dt$ 
can be significantly larger than the usually quoted bounds.}

Therefore we now briefly discuss some aspects of the cosmology of
radion fields.  We will adopt here the same attitude taken
in \cite{ADDlong}.  There, limits were put on the highest
temperature $T_*$ up to which the universe could be considered ``normal",
that is, with the extra dimensions stabilized and energy density
dominated by the radiation on our wall. Since on-the-wall interactions
can produce gravitons which escape into the bulk and which in turn
can variously affect the expansion rate of the universe during
nucleosynthesis, overclose the universe, and unacceptably distort the
background photon spectrum when they decay, the normalcy temperature
$T_*$ was limited to $\sim {\rm few}\mev$ to $\sim 1\gev$ for $n=2-6$.
Fortunately, $T_* \gsim$ 1 MeV in all cases (with $n=2$ marginal), so that
the successful predictions of primordial nucleosynthesis can still hold
in our scenario.  In our present analysis of radion cosmology, we will
be content to show the cosmological safety of the scenario at
temperatures $\lsim T_*$.  Namely, we will assume 
that the early universe at temperatures $\gsim T_*$ evolved into a state
with the radion stabilized, with the energy density stored in radion
oscillations small enough to never overclose the universe.  We will
show that this is enough to guarantee negligible variations in $G_N$, 
and that subsequent interactions with SM fields on the wall will
not significantly excite the radion away from its minimum.\footnote{A
full discussion of the very early universe cosmology in our scenario,
in particular the worry that an early period of inflation could lead
to a form of the Polonyi problem involving $\phi$ will appear in
Ref.~\cite{AHDMRfuture}.}

First, note that since $T_* \ll \tev$, the Hubble expansion rate at all
times of interest satisfies $H \sim T^2/M_{\rm pl} \ll (1{\rm mm})^{-1}$, so
that the expansionary ``friction'' term can not stop $\phi$ from oscillating.
Further, since $\phi$ is so light and gravitationally coupled, it is
essentially stable cosmologically.  The energy density $\rho_{\phi}$ 
stored in its oscillations redshifts away as $1/R^3$, so that $\rho/T^3$
is invariant. It is easy to see that in order for $\phi$ to never
dominate the energy density of the universe, we must have
\be
\frac{\rho_{\phi}}{T^3} \lsim \frac{\rho_{\rm crit.~now}}{T^3_{\rm now}}
\sim 3 \times 10^{-9} \gev.
\ee
Using $\rho_\phi\sim m^2_\phi \phi^2$, this is enough to show
that the variations in $G_N$ are miniscule at all
epochs $T\leq T_*$:
\be
\frac{\delta G_N}{G_N} = n\frac{\delta r}{r_0} \sim \frac{\phi}{M_{\rm pl}}
\sim \frac{\sqrt{\rho_{\phi}}}{m_{\phi} M_{\rm pl}} \lsim 10^{-12}.
\ee
Furthermore, interactions with the SM fields can not significantly
excite the radion into oscillation.  Note that it is only the excitations
of  $\phi$ (and not its associated KK modes) which would correspond to
changing the radius of the extra dimensions (and hence varying $G_N$)
on cosmological scales.  This single mode has couplings suppressed by
the ordinary 4-dimensional Planck scale $\mpl$, and it is therefore
very difficult to excite. Quantitatively, the rate at which collisions of
SM particles dump energy into $\phi$ is
\be
\dot{\rho}_{\phi} \sim {T^7\over M_{\rm pl}^2}.
\ee
The total amount of energy dumped into $\phi$ during a Hubble time
is then 
\be
\delta \rho_{\phi} \sim \frac{T^5}{M_{\rm pl}},
\ee
leading to an unobservably small variation in $G_N$
\be
\frac{\delta G_N}{G_N} \sim
\frac{\sqrt{\delta \rho_{\phi}}}{m_\phi M_{\rm pl}}
\sim \left(\frac{T^5}{(1\tev)^4 \mpl}\right)^{1/2} \lsim 10^{-18}.
\ee
Of course, this is hardly surprising. Recall that at temperatures below
$T_*$, the energy dumped into the bulk gravitons and $\phi$ {\it together
with all of their KK excitations} never overclose the universe.  Even if
all of this energy was somehow transferred into moving the single mode $\phi$, 
we already found that as long as the $\phi$ energy did not overclose
the universe, the variations in $G_N$ are negligible. 

Similar comments apply to radion excitation in stars: the energy lost
to the production of $\phi$ together with all its KK excitations are safe for 
the same reason as bulk graviton production is safe (see Ref.~\cite{ADDlong}),
while the single mode $\phi$ is too weakly coupled to be perturbed enough 
for significant variations of $G_N$ to be observable. For instance, in the
collapse of SN1987A over a time $t_{\rm SN}\sim 1 {\rm s}$, the variation
in $G_N$ can be estimated as above, yielding
\be
{\delta G_N\over G_N}\biggr|_{\rm SN} = n{\delta r\over r_0}
\sim \left({ T_{\rm SN}^{n+7} t_{SN}
\over \mst^{n+2} m^2 \mpl^2}\right)^{1/2} \ll 1,
\ee
even in the worst case $T_{\rm SN}\sim 100\mev$, $n=2$.  Thus 
the local variation of $G_N$ is harmless for any number of extra
dimensions $n=2-6$.

\section{Radius Stabilization Mechanisms}

We now turn to some explicit mechanisms by which internal
dimensions may be stabilized at a radius much greater than the
fundamental Planck length $\sim \mst^{-1}$.
Two issues must be distinguished in discussing radius stabilization:
the mechanism by which the internal dimensions are prevented from
collapsing to $1/\mst$, and the mechanism by which they are prevented
from expanding to a size much larger than a millimeter or fermi.

\subsection{Generalities}

The most obvious idea for limiting the expansion of the 
internal dimensions is to employ a component of the potential
energy that scales like the volume of the internal space:
$V \sim  r^n$.  Such an effective potential energy density
results from a {\it positive} bulk cosmological constant
$\Lambda$, which gives $V(r) \sim \Lambda r^n$  as shown in Section~2. 
As we have already discussed the size of this bulk cosmological
constant must be small.  However, while we have no compelling
explanation for the size of this bulk cosmological constant,
it is interesting that its smallness can at least be stable under
radiative corrections.  Suppose that the short-distance theory of gravity
(perhaps string theory) is supersymmetric, with the supersymmetries broken
only on the walls at a scale $\sim M_* \sim |\bar{f}| \sim 1\tev$.
It is easy to see that the Bose-Fermi splittings induced in the bulk
supergravity multiplet are then \cite{AADD}
\be
|m^2_{\rm bose} - m^2_{\rm fermi}|_{\rm bulk}
\sim N_{\rm wall} \frac{M_*^4}{M^2_{\rm pl}}
\ee
so that the quantum corrections to the bulk cosmological constant
would be of order
\be
|\Lambda_{\rm quant}| \sim  \left(|m^2_{\rm bose} -
m^2_{\rm fermi}|_{\rm bulk} \right)^{(4+n)/2} .
\ee
The ratio of the quantum correction to the tree value is bounded above
once the curvature constraint is used, and we find 
\be
\frac{\Lambda_{\rm quant}}{\Lambda_{\rm tree}}
\lsim \left(\frac{M_*}{M_{\rm pl}}\right)^{(2n + 4)/n} \ll 1.
\ee
Therefore, the small value of the cosmological constant can
be technically natural.

We now turn to the ways in which the radii of the extra dimensions
can be stopped from collapsing to small values.
We will see that a wide range of mechanisms are in principle possible,
leading to a variety of power-law potentials of the form $1/r^\ell$
for various $\ell$.  One minimal possibility is if the compact manifold
has (positive) curvature, in which case
\be
V_{\rm bulk}(r) \sim \Lambda r^n - M_*^{n+2} r^{n-2}
\ee
As can readily be seen from Eq.~(\ref{etaeqn}), this will require a
large positive value for the ratio $\gamma=N_{\rm wall}
\bar{f}^4/\mst^4 $, which can arise if 
we have a configuration of a large number $N_{\rm wall} \gg 1$ of branes.
This possible ``brane lattice crystallization",
together with various generalisations, will be discussed in the next
subsection.  Alternately, if we wish to compactify on manifolds with
no curvature (tori), the ultraviolet stabilization can be provided by
dynamics conserving a topological number $k$, which we will explore in 
Section~4.3.

\subsection{Radius Stabilization from Brane Lattice Crystallization}

The largeness of the internal dimensions compared to
$(1\tev)^{-1}$ can arise from the existence of a large (conserved)
number of branes populating the bulk.  There can exist
inter-brane forces which act like the Van der Walls
and hard-core forces between atoms in a crystal.  The inter-brane
distance is set by these forces, and might be quite small,
but the size of the whole
internal space is set by the total number of branes, just
as the total extent of a crystal is set by the number of atoms,
rather than just the inter-atom distance which is
much smaller.

\begin{description}

\item[{[I]}] {\bf Minimal scenario.}

We will motivate stabilizing
the extra dimensions with large brane numbers by considering the
minimal example of compact manifolds with positive curvature, which 
together with a positive bulk cosmological constant give a bulk
potential of the form
\be
V_{\rm bulk} = \Lambda r^n - M_*^{n+2} r^{n-2}
\ee
For $n=2$, the curvature contribution to the potential is constant
and does not play any role in radius stabilization (although it does
contribute an extra term to the effective 4-d cosmological constant).
For $n>2$, however, this potential has a stable minimum.
From Eq.~(\ref{etaeqn}), we find that a large value for $\gamma$ is needed,
$\gamma \sim (M_* r_0)^{n-2}$.
For simplicity, we will assume that all the branes are broadly similar
with $\bar{f}^4 \sim M_*^4$.
Then, we must have a large number of branes
\be
N_{\rm wall} \sim (M_* r_0)^{n-2} \sim
\left(\frac{M_{\rm pl}}{M_*}\right)^{2(n-2)/n}.
\label{numberbranes}
\ee
Numerically this varies from $N_{\rm wall} \sim 10^{10}$ for $n=3$
to $N_{\rm wall} \sim 10^{20}$ for $n=6$.

Although this is a large number it is not so large as to lead
to problems.  Specifically, we note that
there is a constraint on the total number of branes that can
populate the internal dimensions.  If the transverse inter-brane
separation becomes comparable to $1/\mst$, then there will be new
light open string modes that arise from strings starting
on one brane and ending on a neighbor.  
Thus the maximum number of branes that can occupy the extra
dimensions is
\be
N_{\rm wall,~max} \sim (r_0)^n \mst^n \sim
\left({\mpl\over\mst}\right)^2 \sim 10^{32},
\label{maxbranes}
\ee
which is considerably greater than the necessary number
Eq.~(\ref{numberbranes}).

However, with such a large number of branes, it is obviously
important to ensure that
some dynamics forces them to spread out in the bulk and not sit on
top of each other.  This can easily be arranged.
We know that there is a gravitational force between the branes, and
if they carry any sort of (like sign)
gauge charge there will also be an opposite gauge force between them
with exactly the same dependence on inter-brane separation.
In fact, when the charge density $\rho$ is equal to the tension $T$,
there can be an exact cancellation of the inter-brane forces.
This is what happens in the case of supersymmetric D-branes.
Polchinski's now classic calculation of the forces between D-branes
demonstrated that the forces due to Ramond-Ramond gauge fields
precisely cancelled the gravitational forces in the
supersymmetric limit, as they must 
for a pair of BPS states, which satisfy $T=\rho$. If there is a mismatch
between the charge and tension of the branes, 
the net force between a pair of branes can be made repulsive, forcing
them to spread out uniformly in the bulk.  Of course, we must now take
the inter-brane potential energy into account in the energetics, but
interestingly, this effect is parametrically of the same 
order as the terms in the potential we already have.  By Gauss' law,
the potential between branes falls off with the inter-brane separation
$r$ according to the coulomb potential in the transverse $n$ dimensions
$V_{\rm int.}(r) \sim \rho^2/r^{n-2}$.

If we first imagine just two 3-branes populating the internal space
the potential energy varies as
\be
V(r) \sim \mst^4 {1\over (r\mst)^{n-2}}.
\label{classical} 
\ee
Here we have taken the effective net charge density on the wall
to be $\mst^4$, as we would expect if supersymmetry is broken
at a scale $|f|\sim \mst$. 
The inter-brane distance can be estimated from balancing this
repulsive force against a bulk cosmological constant term
$V(r) \sim {\bar\Lambda} r^n$.  Imposing the cancellation of
the 4-dimensional cosmological constant Eq.~(\ref{IRcosmo}),
leads to an inter-brane separation $r_I$
\be
(r_I)^{n-2} \sim {\mst^4\over \mst^{n+2}} \sim (1\tev)^{(2-n)}.
\label{interbranesep}
\ee

What happens when $\nwall$ branes occupy the internal space?  One may
think that the size of the internal volume will just be $\nwall$
times larger
than $(r_I)^n$ calculated above.  However this is incorrect.  The
reasons for this are two-fold.  The first is that, unlike in a normal
crystal, there is no necessity that the inter-brane forces are screened.
Thus the total potential energy density due to the
inter-brane forces increases as $\nwall^2$, just as the gravitational
potential in a star, and the
UV stabilizing part of the potential has the form
\be
V \sim \mst^4 \nwall^2 {1\over (r\mst)^{n-2}},
\label{classicalQ}
\ee
where $r$ is now roughly the total extent of the system.  The second
reason why the two brane calculation is modified is that the
equation for the cancellation of the effective 4-dimensional IR
cosmological constant is modified to Eq.~(\ref{IRcosmoN}). 
Putting all parts of the potential together, we have
\be
V(r)_{\rm tot} \sim \mst^4\nwall^2 {1\over (r\mst)^{n-2}}
+ \bar \Lambda r^n + \nwall f^4 .
\label{modtotalpot}
\ee
Solving for the size of the system gives,
\be
r_0 \sim \left({\nwall^2 \over {\bar\Lambda}\mst^{n-6}}\right)^{1/2(n-1)}.
\ee
However $\bar\Lambda$ can be eliminated by imposing Eq.~(\ref{IRcosmoN}),
with $|f|\sim\mst$ leading to
\be
\bar \Lambda \sim { \mst^{(n+4)} \over \nwall^{2/(n-2)} },
\label{lambdaQ}
\ee
and thus the final expression for $r_0$
\be
r_0 \sim { \nwall^{1/(n-2)} \over \mst }.
\label{r0Q}
\ee
Utilizing the formula for the required
size of the extra dimensions, $(r_0)^n\mst^{n+2} = \mpl^2$,
we can solve for the necessary brane-number
\be
\nwall \sim \left({ \mpl\over \mst}\right)^{2(n-2)/n} ,
\label{Qnumber}
\ee
exactly the expression Eq.~(\ref{numberbranes}).

Notice that if one substitutes this value back into the equation
for $\bar\Lambda$, Eq.~(\ref{lambdaQ}), then one finds
\be
\bar\Lambda \sim \mst^{n+4} \left({\mst\over \mpl}\right)^{4/n},
\label{lambdaQvalue}
\ee
which is smaller than the naive value $\mst^{n+4}$, showing
that indeed one component of the hierarchy problem in this
framework is the (bulk) cosmological constant problem. 

As discussed above, there is one other
requirement that needs to be satisfied. 
The mean curvature radius on scales smaller than the inter-brane
separation needs to be larger than the inter-brane separation
itself.  The average inter-brane transverse separation is
now
\be
r_I \equiv \left({(r_0)^n \over \nwall}\right)^{1/n} \sim
{1\over \mst} \left({\mpl \over \mst}\right)^{4/n^2},
\label{interbraneQ}
\ee
whilst the curvature radius resulting from our potential is
\be
L_{\rm curv} \sim {1\over \mst} \left({\mpl \over \mst}\right)^{2/n}.
\label{curvQ}
\ee
For the case of $n>2$ where the above analysis applies, one
always has $L_{\rm curv} > r_I$ as required.

In addition if the supersymmetries
of string theory are broken only by on-the-wall dynamics at a scale
$\sim \mst \sim 1\tev$, then the mass splittings so induced among the
bulk supergravity multiplet are $\sim \mst^2/\mpl\sqrt{\nwall}$,
and a bulk cosmological constant of order
$\Lambda\sim(\mst^{2}/\mpl\sqrt{\nwall})^{(4+n)}$ 
arises.  Then the ratio of this new term to the $\bar\Lambda$ term is
\be
{\Lambda_{\rm quant.} \over \Lambda_{\rm tree}}
\sim \left({\mst\over \mpl}\right)^{(2n+4)/n} \ll 1 .
\label{ratioterms}
\ee
Therefore the value of the bulk cosmological constant can still
be technically natural in the case of a large number, $\nwall$,
of branes.

In summary, we have made a number of simplifying assumptions which
can be questioned and modified.  These include the simplification
that all 3-branes are broadly similar
and have tensions and charge densities $\sim(1\tev)^4$
with a mismatch that is also of order $(1\tev)^4$. 
Nevertheless, it is encouraging that the large-brane-number scenario for
stabilizing the volume of the internal dimensions at large values
passes the first tests.

\item[{[II]}]{\bf Non-extensive Bulk Cosmological Constant.}

There is another interesting possibility where the size of the
extra dimensions is completely explained by a large brane number
without needing to invoke another small 
parameter (the small bulk cosmological constant in the above
analysis).  Suppose that the 
the IR potential is $\lambda r^a$, for $a<n$ with the normal
$\Lambda r^n$ term being sub-dominant.  The bulk potential then
reads
\be
V(r)_{\rm bulk} \sim \frac{N_{\rm wall}^2}{M_*^{n-6} r^{n-2}}
+ \lambda r^a
\label{modtotalpot2}
\ee
With this potential we still need the same large brane number
$N_{\rm wall} \sim (M_* r_0)^{n-2}$.
The size of $\lambda$ is to be
\be
\lambda \sim M_*^{4+a}
\left(\frac{M_*}{M_{\rm pl}}\right)^{2(a+2-n)/n}.
\ee
For $a=n-2$, the required value for $\lambda$ agrees with the
natural value $\sim M_*^{4+a}$.  This is intriguing, since $a=n-2$
is precisely the power associated with curvature terms! 
Of course, a compact manifold with positive curvature makes the
wrong-sign contribution to the potential, but we can choose the
compact manifold to have negative curvature.  For instance, genus
$g>1$ Riemann surfaces have negative Euler characteristic and hence
negative average curvature by the Gauss-Bonnet theorem. 
In order to stabilize more than two dimensions in this way, we
can compactify on direct products of such Riemann surfaces, which
will then give the correct exponent and the correct sign in the
potential of Eq.(\ref{modtotalpot2}). 

\item[{[III]}]{\bf Casimir forces between branes.}\footnote{This
possibility was also analyzed by Sundrum~\cite{Raman2}.}

Another potentially attractive idea for UV stabilization at the quantum
level is to use the Casimir force to maintain the size of the
internal space~\cite{CW}.  The effective 4d potential energy density 
corresponding to the Casimir effect in a $(4+n)$-dimensional
spacetime is
\be
V(r) \sim {C\over r^4},
\label{casimir}
\ee
where $C$ is a calculable coefficient in any given model.  Even
with a general non-extensive stabilizing potential,
$\Delta V \sim \lambda r^a$ this leads to a inter-brane
distance of
\be
r_I \sim \left({C\over \lambda}\right)^{1/(4+a)} .
\label{casimirrI}
\ee
Given that the ``natural'' value of $\lambda$ is expected to be
$\mst^{(4+a)}$, this
clearly doesn't allow us to stabilize at large radii.  What about
many branes?  The problem is that, when we go to
$N_{\rm wall}$ 3-branes in the bulk, the Casimir energy does
not increase with $N_{\rm wall}$ for $n\ge 2$.  But the
total wall cosmological constant $N_{\rm wall} \bar{f}^4$ does,
and thus the situation gets worse.

In summary, the Casimir force idea, even with a large brane number
$N_{\rm wall} \gg 1$, fails to stabilize the internal dimensions
at large radii, at least under the simplifying assumptions we have made.

\end{description}

\subsection{Topological Stabilization}

One of the most attractive ways of preventing collapse is to
imagine that there is a topologically conserved quantity which
holds up the size of the extra dimensions.  A prototypical
example of this is provided by the monopole stabilization
mechanisms discussed in Ref.~\cite{monopole}
and in the context of our scheme by Sundrum~\cite{Raman2}.
Consider the simple case of two extra dimensions and where the internal
manifold has the topology of a 2-sphere, $S^2$.  Further suppose that in the
bulk there exists not only the graviton, but also a U(1) gauge field, which
might naturally be a Ramond-Ramond (RR) gauge field of the string theory
in question.
Then it is possible to take the gauge field configuration on $S^2$ to be
topologically non-trivial with quantized ``monopole number'' $k$
(the first Chern number of the U(1) bundle) given
by\footnote{We will always use $H$ for field strengths of gauge fields
that live in the bulk.  Quite often we will think of these as being RR
gauge fields.}
\be
\frac{1}{2\pi} \int_{S^2} H^{(2)} = k .
\label{monopolek}
\ee
If the area of the $S^2$ is denoted $V_{(2)}$ then we have
$H\sim k/V_{(2)}$ and since the kinetic term for the U(1)
gauge field is (expressed in form notation, with $M^4$ denoting
4-dimensional Minkowski space)
\be
S_{KE} \sim \frac{\mst^2}{g^2} \int_{M^4 \times S^2} H\wedge^*H ,
\label{u1kinetic}
\ee
we have that the 4-dimensional potential energy density of the
monopole field on the $S^2$ scales like 
\be
V\sim  \frac{\mst^2}{g^2} {k^2 \over V_{(2)} } .
\label{u1potential}
\ee
In other words we get an energy density that scales
like $k^2 \mst^2/(g^2 r^2)$.  For large enough monopole
number, $k$, this will stabilize the internal $S^2$
at any desired size.

This basic mechanism has a wide variety of generalizations.  One such
is to use the topological invariants of the higher-form RR
gauge fields that naturally arise in the type II and type I string
theories with D-branes.  Let us recall here that for
stabilizing $n>2$ dimensions topologically, 
we must work with compact manifolds of zero curvature i.e. tori, 
since otherwise the curvature term will dominate the dynamics
and we must revert to the analysis of the previous subsection.

\begin{description}

\item[{[I]}]{\bf Higher-form RR fields}

Denote the manifold
of the extra $n$ dimensions by $E^n$, and suppose that the bulk
theory contains an $(n-1)$-form U(1) gauge field, with $n$-form field
strength $F^{(n)}$.  Then once again there is the topological invariant
\be
\frac{\mst^{n-2}}{2\pi}\int_{E^n} H^{(n)} = k .
\ee 
The kinetic energy of $H^{(n)}$ is the generalization of
the usual 1-form gauge kinetic term
\be
S_{KE,n} \sim \frac{\mst^n}{g^2} \int_{M^4 \times E^n}
H^{(n)}\wedge^*H^{(n)} ,
\label{nformkinetic}
\ee
and thus the potential energy density depends on the volume $V_{(n)}$
of $E^n$ as 
\be
V\sim  \frac{1}{g^2 \mst^{n-4}} {k^2 \over V_{(n)} } .
\label{nformpot}
\ee

In the case of the chiral type IIB string theory there exists
1 and 3-form RR field strengths and a self dual 5-form RR field strength
(together, of course, with their magnetic duals).
There also exists the usual NS-NS 3-form field
strength.  The type I string
theory has a 3-form RR field strength and it's 7-form magnetic dual.
Thus using the invariants so far described, it is natural to stabilize
1, 3, and 5-manifolds. 

However, invariants that lead to $1/r^n$ potentials for $E^n$ are not
the only possibility.  Consider the situation in which our 3-brane
world is the boundary of (a set of) higher-dimensional branes
which are in turn embedded in the full $(4+n)$-dimensional space.
We can then use topological invariants of the world-volume gauge
fields of these higher-dimensional branes to stabilize the
internal dimensions.  To make this clear consider the following
very simple example: In the $n=4$ case take the internal manifold
to be $E^4 = T^2_1 \times T^2_2$.  Suppose further that there
exist 2 5-branes that intersect at the position of our 3-brane
but are perpendicular in the extra 4 dimensions, so that
one 5-brane lives in $M^4\times T^2_1$ and the second lives in
$M^4\times T^2_2$.  Then we have the two topological invariants
\be
\frac{1}{2\pi}\int_{T^2_i} F_i = k_i, \qquad i=1,2,
\label{twotopolog}
\ee
where $F_i$, $i=1,2$ are world-volume U(1) 2-form field strengths
of the first and second 5-brane.  The brane-localized kinetic terms
for these gauge fields then leads to an effective 4-dimensional
potential energy density of the form
\be
\Delta V(r) \sim {\mst^2 k_1^2 \over r_1^2} +
{\mst^2 k_2^2 \over r_2^2},
\label{two5branes}
\ee
where $r_1$ and $r_2$ are the radii of the two $T^2$'s.  
Note that since we have used tori, there is no negative
curvature term $\sim -M_*^2 r^2$ in the potential.
This, then, is an UV stabilizing potential for
$E^4= T^2_1 \times T^2_2$ not of the form $1/r^4$.
Clearly this type of mechanism admits many generalizations.

Finally, one can also consider higher ``reducible''
invariants such as the second Chern class of the usual 2-form U(1)
field strength defined wrt a 4-manifold
\be
\frac{1}{8\pi^2}\int_{E^4} H^{(2)}\wedge H^{(2)} = c_2, 
\ee 
but such invariants typically lead to a potential energy varying
as $r^\alpha$ with $\alpha\ge 0$.

\item[{[II]}]{\bf Metric topological invariants.}

Purely metric topological invariants are possible, for example
the Euler number of a 2-manifold component $E^2$ of the internal
space  
\be
\chi = \frac{1}{2\pi}\int_{E^2} R,
\label{euler}
\ee
where $R$ is the curvature 2-form.
Other possibilities include the Pontrjagin classes of the
tangent bundle of the internal manifold.
However, because the leading term in the gravitational
effective action is only linear in the curvature, this
does not provide a UV stabilizing potential unless higher
derivative terms, such as
\be
\Delta S \sim \mst^n \int {\rm tr}(R^2)  
\label{higherR }
\ee 
are included in the effective action.  For the simple case
of $n=2$ this leads to a potential $V\sim \chi^2 \mst^2/r^2$.
For this to balance, at the appropriate $r_0$, even a
best-case stabilizing potential of the form $\mst^5 r$,
an Euler number of
$\chi\sim (\mpl/\mst)^{3/2}$ is required.  So clearly the 
internal manifold is very highly curved.  In particular, the
leading gravitational action $\mst^4\int {\cal R}$ dominates
the other terms by an amount $(\mpl/\mst)^{1/2}$, and leads
to an unacceptably large bulk cosmological constant.  This seems to
be a generic problem with this type of topological stabilization,
although we have not investigated the question in detail.

\item[{[III]}]{\bf Scalar-field and other non-gauge invariants.}\footnote{Gia
Dvali has independently considered this possibility.  We thank him for
discussions.}

One can also imagine stabilizing the size of the internal space by the
use of non-gauge or metric topological invariants.  For example,
consider a complex scalar field that lives
on a 1-dimensional higher brane that has as boundary
our 3-brane.  Then the phase of this field can wind as an
$S^1$ cycle of the internal space is transversed, with
topologically conserved winding number 
\be
k = \int_{S^1} d\phi.
\ee
Once again the kinetic
energy of this configuration increases as the size of the internal
space is reduced, and thus a stabilizing potential results.
In order to stabilize more than one dimension in this way, we can have
$n$ different scalar fields living on $n$ different 4-branes which have
the 3 dimensions of our 3-brane in common but have mutually orthogonal 
fourth spatial dimensions. The $i$'th scalar field can wrap around the
4th dimension of the $i$'th 4-brane, generating a potential of the form
$r_1^{-1} + \cdots r_n^{-1}$. Together with a bulk cosmological constant
giving a potential $\sim (r_1 \cdots r_n)$, this can stabilize all the 
$n$ dimensions. 
More sophisticated scalar field invariants are also conceivable, the
Hopf winding number of the map $\pi: S^3 \rightarrow S^2$ being one
among many such examples.  
In general scalar field invariants lead to quite similar
results to the gauge field topological stabilization mechanisms,
but possibly without the natural advantage of gauge fields 
of their constrained couplings.  (For instance it is easy to arrange
that the stabilizing gauge fields do not lead to dangerous flavor-changing
neutral current processes on the wall, while this requires additional
input in the scalar case.)

\item[{[IV]}]{\bf Phenomenologically successful topological stabilization.}

In the previous subsections we have seen that a variety of UV
stabilizing potential energy densities of the general form
$M_*^4(k^2/(r M_*)^\beta)$ are possible. 
We now wish to get an idea of the numerical values the various
parameters must take to stabilize the radii at
the desired sizes.  From Eq.~(\ref{Neqn}), we have that
\be
k^2 \sim \frac{|N_{\rm wall} \bar{f}^4|}{M_*^4} (\mst r_0)^\beta
\gsim \left({\mpl\over
\mst}\right)^{2\beta/n} \left({1\tev\over \mst}\right)^4.
\label{kbound}
\ee
where the inequality comes from imposing the constraint that
the radion is heavier than $\sim (1$mm$)^{-1}$.
In the above examples of topological radius stabilization the quantity
$k$ is directly proportional to the ``monopole'' number.
From this expression
the smallest $k$ clearly occurs when the ratio $\beta/n$ is as small
as possible.  As an example, if $\beta=1$ and there 6 extra dimensions,
then for $M_* \sim 1\tev$, $k\gsim 3\times 10^2$ is required
to stabilize at a sufficiently
large radius.  If $\beta=2$ and there are 6 extra dimensions, then
$k\gsim 10^5$ is necessary.  The $\beta=2$ case is particularly
interesting since it is the first case we can realize with
gauge-field topological invariants rather than scalar field
invariants.  The worst case, requiring the largest $k$, occurs when
$\beta/n$ takes on its largest value.  A typical ``worst-case''
is provided by the irreducible topological stabilization mechanism
involving bulk RR fields (for example).  This gives $\beta= n$, and
leads to $k\gsim 10^{15}$. 

Note that in the special case with $\beta = 1$ and $n=6$, if we are
willing to move the fundamental scale $M_*$ up to $\sim 10\tev$,
while keeping $N_{\rm wall} \bar{f}^4 \sim (1\tev)^4$,
we can get away with $k \sim 1$.  This may not be unnatural, after
all, one could easily imagine that the scale
$\bar{f}$ is $\sim 10$ smaller than the scale $M_*$ due to 
partial cancellations up to some 1 or 2-loop order. 
Also recall that as shown in Section~4.1, the small value of
the bulk cosmological constant is unexplained but is at least
technically natural, if SUSY is primordially broken on the walls.

In summary we have shown that the topological stabilization
mechanism successfully meets all our phenomenological
requirements, with a price of a large, but in some cases
not too large integer $k$.

\item[{[V]}]{\bf Corrections to leading-order potentials.}

Finally, one may worry that in the regime of interest, when $r\sim r_0$,
the semiclassical reasoning that we have applied to the leading-order
kinetic and non-derivative terms in the effective action suffers from
large corrections due to the presence of other terms.  Such corrections
are, in actual fact, entirely negligible.  For example, 
if one included higher-order derivative terms, such
as 
\be
\Delta S \sim \mst^{n-4} \int_{M^4 \times E^n}
(H^{(p)}\wedge H^{(p)})\wedge^*(H^{(p)}\wedge H^{(p)}) ,
\label{higherkinetic}
\ee
in the effective action, then they would lead to corrections in the
4-dimensional effective potential energy density, $V$, of order
\be
\Delta V \sim \mst^4 { k^4 \over (\mst r_0)^{3n}}
\ee
at the minimum $r_0$.  Compared to the leading kinetic term
this is a fractional change of order 
\be 
{\Delta V \over V} \sim k^2 \left({\mst\over \mpl}\right)^4 ,
\ee
negligible unless $k\gsim 10^{30}$.  Such statements generally
apply for $r\sim r_0$, and are basically due the fact that
$r_0 \gg (1\tev)^{-1}$.
This is not quite trivial because of the potentially large
dimensionless factor $k$ which could have overcome this
suppression.  In any case we see that the leading-order
analysis is entirely sufficient unless we are interested in
physics at radii $r \ll r_0$.

\end{description}

\section{Remarks and Summary}
   
The hierarchy problem in our framework is replaced by the problem of  
obtaining large new dimensions, of a size which varies between a
millimeter and a fermi depending of the number of new dimensions, in  
a theory with a much smaller fundamental length $\sim \tev^{-1}$.  
In this paper we exhibited mechanisms which provide such large   
extra dimensions.  These mechanisms relied on two ingredients:\\

\begin{itemize}
\item
A large conserved integer $Q$, which can be a large number $N_{\rm wall}$
of branes, or the topological charge $k$ of the vacuum configuration.
This large integer should be regarded as analogous to the net conserved
baryon number which accounts for the large size of macroscopic
objects relative to that of atoms.  The necessity for such a large number
was {\it not} forced on us by the need for large internal dimensions,
but rather by the requirement that the radial oscillation field (or
``radion'') be sufficiently heavy to have escaped tests of gravity
at the millimeter-scale and above.  The value of $Q$ depends
on the details of the stabilization scenario; it varies from
$N_{\rm wall}\sim 10^{10}$ to $N_{\rm wall}\sim 10^{20}$ in the
brane-lattice-crystallization
scenario, while in the topological stabilization scenario it varies
from $k\sim 1$ to $k\sim10^{15}$.
\item
A small bulk cosmological constant, analogous to the 4-dimensional
cosmological constant whose smallness accounts for the size of our
universe relative to the Planck length.  However, as we discuss in
detail in, for example, Section~4.1, the value of this bulk
cosmological constant is {\it stable against radiative
corrections} if supersymmetry-breaking of order the fundamental
Planck mass $\sim\mst\sim 1\tev$ takes place on the 3-branes.
Of course we must still impose a fine tuning to get a vanishing
effective 4-dimensional, brane-localized cosmological constant
in the IR in our world.  This is expressed in Eq.~(\ref{IRcosmo})
or (\ref{IRcosmoN}), depending on the stabilization scenario.
\end{itemize}

A valid criticism of our analysis is that we have not provided
a {\it dynamical} framework in which, for instance, the largeness
of $Q$ or $k$ is explained.  As discussed in the introduction our
viewpoint on this issue is that this is closely analogous to the
situation in the MSSM where soft supersymmetry-breaking operators
of order $(1\tev)$ are introduced~\cite{dimopoulosgeorgi81}.

With the advent of many quantum-field-theoretic (QFT)
models of dynamical supersymmetry breaking
it is commonly believed that the problem of the size of these
soft operators has been solved, at least in principle.  However,
from a fundamental vantage-point this belief is not correct.
Concretely, what is the situation in the standard model
or MSSM, where the usual (reduced) Planck mass
$\mpl\sim 2\times 10^{18}\gev$ is taken as fundamental?  We must
now explain the ratio of this Planck scale to the  
weak scale $\sim 10^{15}$.  There too the ``dilaton runaway problem"  
prevents us from having a calculational framework for this   
number.  This point is important to emphasize.  Although in the context
of QFT dynamical SUSY breaking solves the hierarchy problem, in that
it generates the small scale by dimensional transmutation,
in the context of string theory the couplings and thus the scale of
SUSY breaking are dynamical, and there is a ground state 
at zero coupling with unbroken supersymmetry~\cite{dineseiberg}. 
This means that there exists no known solution to the hierarchy
problem in usual 4-dimensional QFT once it is embedded in string
theory.  Therefore both frameworks face similar challenges.

\section*{Acknowledgements}
It is a pleasure to thank Philip Argyres, Gia Dvali, Michael Graesser,
Vadim Kaplunovsky, Yaron Oz, Riccardo Rattatzi, Eva Silverstein
and Raman Sundrum for valuable discussions.
SD thanks the CERN theory group for its
hospitality.  The work of NAH is supported by
the Department of Energy under contract 
DE-AC03-76SF00515. The work of 
SD is supported in part  
by NSF grant PHY-9219345-004.  The work of JMR is
supported in part by an A.P. Sloan Foundation Fellowship.

\def\pl#1#2#3{{\it Phys. Lett. }{\bf B#1~}(19#2)~#3}
\def\zp#1#2#3{{\it Z. Phys. }{\bf C#1~}(19#2)~#3}
\def\prl#1#2#3{{\it Phys. Rev. Lett. }{\bf #1~}(19#2)~#3}
\def\rmp#1#2#3{{\it Rev. Mod. Phys. }{\bf #1~}(19#2)~#3}
\def\prep#1#2#3{{\it Phys. Rep. }{\bf #1~}(19#2)~#3}
\def\pr#1#2#3{{\it Phys. Rev. }{\bf D#1~}(19#2)~#3}
\def\np#1#2#3{{\it Nucl. Phys. }{\bf B#1~}(19#2)~#3}
\def\mpl#1#2#3{{\it Mod. Phys. Lett. }{\bf #1~}(19#2)~#3}
\def\arnps#1#2#3{{\it Annu. Rev. Nucl. Part. Sci. }{\bf #1~}(19#2)~#3}
\def\sjnp#1#2#3{{\it Sov. J. Nucl. Phys. }{\bf #1~}(19#2)~#3}
\def\jetp#1#2#3{{\it JETP Lett. }{\bf #1~}(19#2)~#3}
\def\app#1#2#3{{\it Acta Phys. Polon. }{\bf #1~}(19#2)~#3}
\def\rnc#1#2#3{{\it Riv. Nuovo Cim. }{\bf #1~}(19#2)~#3}
\def\ap#1#2#3{{\it Ann. Phys. }{\bf #1~}(19#2)~#3}
\def\ptp#1#2#3{{\it Prog. Theor. Phys. }{\bf #1~}(19#2)~#3}


\begin{thebibliography}{99}

\bibitem{ADD}
N. Arkani-Hamed, S. Dimopoulos and G. Dvali, hep-ph/9803315, to appear  
in {\it Phys. Lett.} {\bf B}.

\bibitem{AADD}
I. Antoniadis, N. Arkani-Hamed, S. Dimopoulos and G. Dvali,  
hep-ph/9804398, to appear in {\it Phys. Lett.} {\bf B}.

\bibitem{ADDlong}
N. Arkani-Hamed, S. Dimopoulos and G. Dvali, hep-ph/9807344.

\bibitem{Dbrane} See for example: J. Polchinski, {\it TASI
lectures on D-branes}, hep-th/9611050;\\
C. Bachas, {\it Lectures on D-branes}, hep-th/9806199.

\bibitem{DDG}
K.R. Dienes, E. Dudas and T. Gherghetta, hep-ph/9803466 and
hep-ph/9806292; K.R. Dienes, E. Dudas,T. Gherghetta and
A. Riotto, hep-ph/9809406.

\bibitem{antoniadis}
I. Antoniadis, \pl{246}{90}{377}. 

\bibitem{HW}
P. Horava and E. Witten, \np{460}{96}{506}; E. Witten.\\
J.D. Lykken, \pr{54}{96}{3693};\\
E. Caceres, V.S. Kaplunovsky, I.M.Mandelberg, \np{493}{97}{73}.

\bibitem{tye}
G. Shiu and S.H.H.Tye, hep-th/9805157. 

\bibitem{bachas}
C. Bachas, hep-ph/9807415.
 
\bibitem{Raman1}
R. Sundrum, hep-ph/9805471.

\bibitem{Raman2}
R. Sundrum, hep-ph/9807348.

\bibitem{ADMR}
P.~Argyres, S.~Dimopoulos and J.~March-Russell, hep-th/9808138;
to appear in {\it Phys. Lett.} {\bf B}.

\bibitem{dimopoulosgeorgi81}
S. Dimopoulos and H. Georgi, \np{193}{81}{150}.

\bibitem{CW}
P. Candelas and S. Weinberg, \np{237}{84}{397}.

\bibitem{monopole}
Y. Okada, \pl{150}{85}{103}.

\bibitem{AHDMRfuture}
N. Arkani-Hamed, S. Dimopoulos and J.~March-Russell,
in preparation.

\bibitem{dineseiberg}
M. Dine and N. Seiberg, \pl{162}{85}{299}.

\end{thebibliography}
\end{document}